\documentclass[showpacs,preprintnumbers,amsmath,amssymb]{revtex4}

\usepackage{graphicx}
\usepackage{dcolumn}
\usepackage{bm}

\begin{document}

\title{Tunneling anisotropic magnetoresistance in organic spin-valves}

\author{M. Gr\"{u}newald$^{1,2}$}%
\author{M. Wahler$^{1,2}$}
\author{M. Michelfeit$^{1,2}$}%
\author{C. Gould$^{1,2}$}%
\author{R. Schmidt$^{2,3}$}%
\author{F. W\"{u}rthner$^{2,3}$}%
\author{G. Schmidt$^{1,2,4}$}%
\author{L.W. Molenkamp$^{1,2}$}%

\affiliation{%
$^1$Physikalisches Institut (EP3), $^2$R\"{o}ntgen Center for Complex Material Systems, $^3$Institut f\"{u}r Organische Chemie, Universit\"{a}t W\"{u}rzburg, Am Hubland, 97074 W\"{u}rzburg, Germany \\
$^4$Institut f\"{u}r Physik, Martin-Luther-Universit\"{a}t Halle-Wittenberg, D-06099, Germany* \\
*present and permanent address G.S.\\
Correspondence to G. Schmidt email: georg.schmidt@physik.uni-halle.de}

\begin{abstract}
We report the observation of tunneling anisotropic magnetoresistance (TAMR) in an organic spin-valve-like structure with only one ferromagnetic electrode. The device is based on a new high mobility
perylene diimide-based n-type organic semiconductor. The effect originates from the tunneling injection from the LSMO contact and can thus occur even for organic layers which are too thick to support the assumption of tunneling through the layer. Magnetoresistance measurements show a clear spin-valve signal, with the typical two step switching pattern caused by the magnetocrystalline anisotropy of the epitaxial magnetic electrode.

\end{abstract}

\maketitle

Over the past years a number of spin-valves based on various organic semiconductors and contact
materials have been demonstrated (e.g. \cite{Dediu2002,Xio2004,Maj2006,Xu2007,Wan2005,San2007,Vardeny2010,Sch2009}).
Although some experiments indicate injection of spin polarized carriers\cite{Vardeny2010} and some clearly show tunneling \cite{San2007,Sch2009}, it is still unclear for a number of other results whether their data show tunneling magnetoresistance (TMR) or actual spin injection and consequently giant magnetoresistance (GMR).
Moreover, recent results\cite{Wohlgenannt2010} suggest that spin-valve effects are only observed in devices whose resistance only slightly changes during cooldown. This behavior is typical for devices operated in the tunneling regime, however, it is less common for electron transport in polycristalline or amorphous organic semiconductors.

Nonetheless, both GMR and TMR are related to the switching of a pair of magnetic electrodes between parallel and antiparallel magnetization. For both effects at least some spin conservation throughout the whole thickness of the organic interlayer is necessary. It is, however, known that similar magnetoresistance traces can also be caused by charge carrier injection from a single magnetic electrode into another material by a tunneling process (like charge carrier injection into an OSC), if the magnetic electrode exhibits a suitable magnetocrystalline anisotropy. In that case  tunneling anisotropic magnetoresistance (TAMR) can appear.
\cite{Gou2004}. TAMR can cause the magnetic switching characteristics of the electrodes to modulate the tunneling resistance of the injection contact. This effect is much larger than the well known anisotropic magnetoresistance observed in the electrode itself and can erroneously be identified as a true spin-valve signal, unless suitable control measurements are carried out.

We have fabricated a number of different spin-valve like structures (Fig. \ref{FigSample}a) using the organic semiconductor (OSC) \textit{N,N'}-bis(n-heptafluorobutyl)-3,4:9,10-perylene tetracarboxylic diimide (PTCDI-C4F7, Fig.
\ref{FigSample}b) and three different combinations of magnetic and non-magnetic contact materials.
While most spin-valves reported until now are either based on AlQ$_3$\cite{Xio2004}, which is an amorphous low-mobility n-type semiconductor, or on p-type semiconductors such as P3HT\cite{Maj2006} or TPP\cite{Xu2007}, PTCDI-C4F7 is a new air-stable, high-mobility n-type semiconductor\cite{Oh2007}  from the perylene diimide/dianhydride family whose
most common member is PTCDA. Owing to its favorable electron transport properties, PTCDI-C4F7 should be well-suited for the demonstration of spin-polarized electron transport. Three different device types were investigated, distinguished by the combinations of contact materials as listed in table \ref{Table}.

\begin{table*}[ht]
\begin{tabular}{p{3cm}|p{4cm}|p{3cm}|p{3cm}}

\textit{device} &  \textit{bottom contact} & \textit{PTCDI-C4F7 layer} &
\textit{top contact}   \\
\hline \hline
1 &      $15\,nm$ LSMO  &      $100\,nm$&      $30\,nm$ CoFe\\
\hline
2 &      $10\,nm$ Al  &      $250\,nm$&      $30\,nm$ Al\\
\hline
3 &      $10\,nm$ LSMO  &      $150\,nm$&      $15\,nm$ Al\\

\end{tabular}
\caption{Contact materials and layer thicknesses of the three investigated multilayers.}
\label{Table}
\end{table*}

Type-1 devices which have an LSMO bottom- and a ferromagnetic metal top-contact are similar to
those reported for most organic spin-valves (OSVs) in the literature and serve to demonstrate the
suitability of PTCDI-C4F7 for spin-valve operation. Type-2 devices are control samples with only
non-magnetic electrodes and help to exclude possible artifacts caused by the OSC and the well known Organic magnetoresistance (OMAR)\cite{Fra2004} which is also observed in completely non-magnetic layer stacks. Type-3 devices
have one magnetic LSMO electrode and a non-magnetic Al counter electrode. These samples are
intended for investigations on the occurrence of tunneling anisotropic magnetoresistance (TAMR). All
layer stacks are deposited in a UHV chamber designed to allow for the deposition of OSC and metal
layers in direct sequence without breaking the UHV.

The LSMO bottom contacts are fabricated from 10 or 15 nm thick LSMO layers, grown by pulsed
plasma deposition\cite{Ber2004} on Strontium Titanate substrates. For device fabrication, first Ti/Au metal
stripes are deposited on the LSMO, using optical lithography and lift-off. These stripes serve as
alignment marks and later as bondpads. A rectangular bottom contact is then patterned into the LSMO
layer by optical lithography and dry etching, leaving the metal contact at one side of the
rectangle. Subsequently, the sample is inserted into the UHV-deposition chamber where a bake-out
procedure is performed at 450 $^\circ$C for 1 hour at an oxygen pressure of 10$^{-5}$ mbar, in
order to compensate under-oxygenation which may occur during the processing. Subsequently, the
PTCDI-C4F7 layer and the metal top electrode are deposited under different angles of incidence
through a shadow mask with a rectangular opening. After removing the sample from the UHV chamber,
Ti/Au stripes are deposited through a second shadow mask with striped windows. These metal stripes
are later used as bond pads for the top contacts and also serve as an etch mask for the removal of the top electrode material between the stripes by
dry etching. Samples with aluminum bottom
electrode are fabricated on a Si substrate with a 200 nm thick thermal SiO$_2$ cover layer. A Ti/Au
contact pad is deposited before inserting the sample into the UHV chamber where the bottom aluminum
layer, the PTCDI-C4F7 layer and the top aluminum layer are evaporated through the shadow mask at
three different angles of incidence. The different angles are necessary in order to allow for
insulation between top and bottom electrode. After removing the sample from the UHV chamber the
processing continues in the same way as for the LSMO based samples. This approach provides clean,
oxygen-free, and reproducible interfaces. The samples are characterized at various temperatures
between 4.2 K and room temperature, either in a flow cryostate with an external room temperature
electromagnet (600 mT) or in a $^4$He bath cryostate with a vector field magnet in which magnetic
fields up to 400 mT can be applied in any direction. In the measurements all three samples show less than one order of magnitude increase of resistance during cooldown from room temperature to 4.2 K.

A typical magnetoresistance trace of a type I sample 4.2 K is shown in Fig. \ref{FigMR30}a. The B field is applied along the long axis of the stripe-like device. The magnetoresistance has at least two distinct components. The first comprises the two switching events for each of the scan directions, which are usually attributed to spin-valve operation. For this device this effect is negative (as often described in literature e.g. \cite{Xio2004, Xu2007, Wan2005}) and its size is approx. 8\%. The other component is a continuous increase of the resistance with increasing magnetic field also observed in various experiments \cite{Vin2008, Xu2007, Maj2006-2, Wan2007} which may be attributed to the magnetic saturation of the electrodes. When the B-field is again applied in the plane but perpendicular to the stripe (Fig. \ref{FigMR30}b) the shape of the curve changes. This change in shape can be explained by shape anisotropy of the electrodes leading to a rotation rather than switching. Although the zero field resistance of both measurements is identical, it should be noted that the resistance at 350 mT is different, indicating traces of TAMR.

The type-2 layer stack, which has no magnetic electrodes, is used to identify any magnetoresistance in the pure OSC which might be misinterpreted as a spin-valve signal. The type-2 devices exhibit no detectable magnetoresistance, neither spin-valve nor OMAR (between -400 mT and +400 mT, with an error of less than $\pm 0.05\%$). We can thus exclude OSC-related magnetoresistance effects as explanation for the effects found in samples with magnetic contact layers.

The 3rd type of layer stack, with a ferromagnetic LSMO bottom contact and a non-magnetic aluminum
counter electrode, has an OSC layer thickness of 150 nm. Given the thickness of the OSC layer we can exclude any transport by direct tunneling through the OSC. Even multi step tunneling which has recently been discussed \cite{Sch2009} can not explain the observed transport phenomena. Nevertheless, the high resistance at room temperature and the weak temperature dependence (increase in resistance by one order of magnitude between room temperature and 4.2 K) indicates that the I/V characteristics are dominated by the charge injection processes at the metal/OSC Schottky contacts\cite{For2001}, explaining why the I/V characteristiscs we observe (insert Fig. \ref{Fig0_90_MR}) are strongly reminiscent of tunneling processes.

Because there is only one magnetic layer present in type-3 stacks, no genuine spin-valve signal can
be expected. Nevertheless, the magnetoresistance scans show the two-state behaviour characteristics
of a spin-valve with two ferromagnetic layers. The magnetoresistance curves taken at a bias voltage of 255 mV (Fig.
\ref{FigWaterfall}) clearly exhibit two spin-valve like switching events superimposed to the background. Both, background and switching are well known from organic spinvalves. In this structure, however, they can neither originate from GMR nor from TMR, both TMR and GMR requiring two
ferromagnetic electrodes.

As TMR and GMR as well as any intrinsic magnetoresistance of the LSMO layer\cite{AMRnote} can be excluded here, we interpret these data as tunneling anisotropic magnetoresistance (TAMR), an effect which was first observed in (Ga,Mn)As \cite{Gou2004} and has since been reported to occur in various tunnel contacts on ferromagnetic semiconductors and metals\cite{Rus2005,Bol2006,Mos2007,Wun2008}. This effect allows for spin-valve
functionality in layer stacks with only one ferromagnetic contact.

TAMR originates from spin orbit coupling in a ferromagnetic electrode with crystalline anisotropy. In these electrodes spin orbit coupling can translate any change of the magnetization vector into a change in the density of states (DOS). The k-dependence of the DOS in the electrode has a (weak) dependence on the relative orientation of the magnetization vector with respect to the crystal axes. If the electrode is part of a tunneling contact, the DOS component of the tunneling matrix element of any state strongly depends on the k-vector of the respective state and thus the change in magnetization is reflected in the tunneling resistance. The magnitude of the effect increases with increasing anisotropy and increasing spin orbit coupling.

The typical spin-valve like signature with two pronounced switching events can appear for example in systems with a biaxial magnetocrystalline anisotropy. In such a system a magnetic field sweep at an angle between the two easy axes usually leads to a magnetization reversal via a two step switching process combined with small rotations \cite{Cow1995}. Here, symmetry breaking for example from the growth strain can cause the tunneling matrix element to be different for magnetization alignment along the two different easy axes. In this case the system has four stable magnetization states with two different tunneling resistance values (one for each easy axis) which we may call R1 and R2. The two step switching process then leads either from R1 to R2 and back or from R2 to R1 and back. Depending on the starting point of the magnetoresistance scan the magnetoresistance trace corresponds to a positive or a negative spin-valve effect. The respective switching fields which are observed depend on the field angle with respect to the easy axes. Only for a magnetic field applied exactly along one of the easy axes single step switching occurs and the spin-valve effect vanishes. It is noteworthy that in many (however, not all) organic spin-valves with LSMO electrodes in literature a negative spin-valve effect is observed.

Obviously this phenomenology can be used to identify TAMR. We have performed magnetoresistance scans with different directions of the magnetic field in the plane. In these scans, the coercive fields of the two switching events ($H_{C1}$ and $H_{C2}$) vary depending on the relative orientation of the magnetic
field with respect to the easy axes as expected. Experimental data obtained on our type-3 sample are shown in Fig. \ref{FigWaterfall} (0$^\circ$ and 90$^\circ$ are the two sample diagonals, see also Fig.\ref{FigSample}).

The effect is even clearer in a magnetoresistance plot where the two scans for field orientations of 0$^\circ$ and 90$^\circ$ are shown together on the same scale (Fig. \ref{Fig0_90_MR}). For 90$^\circ$ the resistance is in the high-state in magnetic saturation\cite{Saturationnote}, and it is low between $H_{C1}$ and $H_{C2}$, while for 0$^\circ$ orientation the resistance in saturation is in its low-state and it is high in between H$_{C1}$ and H$_{C2}$. These two directions are the
main axes which determine the minimum and maximum value of the tunneling resistance. The
observation that the maximum resistance value between H$_{C1}$ and H$_{C2}$ (0$^\circ$ curve) is
higher than the maximum resistance in saturation (90$^\circ$ curve) can be explained by the
occurrence of multi-domain behavior in which parts of the sample are magnetized out of plane due to
a weak out-of-plane easy axis. SQUID measurements with the B-field perpendicular to the surface indeed show a
remanent magnetization component.

Still further information comes from a magnetoresistance scan on the type-3 sample in which the
direction of the B-field is rotated by 360$^\circ$ in the plane while its magnitude is kept
constant close to saturation\cite{Saturationnote} (full saturation is not possible in our magnet). For ordinary TMR or GMR with
two ferromagnetic electrodes this scan must always show the same resistance value because the two
layers are always aligned in parallel. For our sample we see an anisotropic resistance distribution
(Fig. \ref{FigPHI}) with a biaxial symmetry, clearly indicating the presence of TAMR.

It should be noted that in order for TAMR to be observed, electron tunneling through the organic layer is not necessary. It is sufficient to have carrier injection into the OSC through a barrier (as in most OSV) and thus to have a tunneling contribution in the injection process. For a sufficiently high contact resistance, the TAMR will always be visible above the normal device resistance, especially at low bias voltages. If however, the resistance of the semiconductor massively increases at low temperatures the effect will no longer be detectable.

We have thus demonstrated that TAMR can exist in OSV structures with at least one LSMO electrode.
This effect opens up new perspective for organic spintronics. However, as the effect can be either
positive or negative in sign, depending on the direction of the applied magnetic field, our
observations imply that careful measurements on any OSV are necessary in order to distinguish
between TAMR and real spin-valve operation.

We thank the EU for funding the research in the project OFSPIN.

\clearpage

\begin{figure}
\caption{Schematic drawing of the vertical transport structure (a). A is the bottom contact material and B is
the top contact material as listed in table 1 for the respective samples. (b) Structure of the PTCDI-C4F7
molecule (b).\label{FigSample}}
\end{figure}

\begin{figure}
\caption{Magnetoresistance trace of a PTCDI-C4F7-based vertical OSV structure with an LSMO bottom electrode
and a CoFe top electrode (type-1 sample). Applying the B-field along the stripe (a) yields pronounced switching events while a measurement with B-field perpendicular to the stripe (b) indicates a rotation of the stripes magnetization.\label{FigMR30}}
\end{figure}

\begin{figure}
\caption{Magnetoresistance sweeps for the TAMR test sample with the magnetic field applied in different
directions in the plane. After each scan the field direction is rotated by 30 $^\circ$. 0$^\circ$ and
90$^\circ$ are the sample diagonals. For 90$^\circ$ and 270$^\circ$ the switching event is towards negative
values. For clarity, scans for different directions are offset by 850 k$\Omega$.\label{FigWaterfall}}
\end{figure}

\begin{figure}
\caption{Magnetoresistance traces (type-3 sample) with the magnetic field aligned along 0$^\circ$ and
90$^\circ$. At large positive or negative fields two different resistance states can be distinguished, a
clear signature of TAMR. For the scan in 0$^\circ$ direction, the spin-valve signal is positive while it is
negative for the 90$^\circ$ sweep. The insert shows the I/V-characteristics of the type-3 device at room temperature and at 4.2 K.\label{Fig0_90_MR}}
\end{figure}

\begin{figure}
\caption{Magnetoresistance scan (type-3 sample) taken at constant field while the angle of the applied field
is slowly rotated by 360$^\circ$. The scan clearly shows the minimum and maximum resistance state at
$\Phi$=0$^\circ$/180$^\circ$ and $\Phi$=90$^\circ$/270$^\circ$, respectively, corresponding to the saturation
states in Fig. \ref{Fig0_90_MR}.\label{FigPHI}}
\end{figure}

\clearpage
\begin{figure}
\includegraphics[width=18cm]{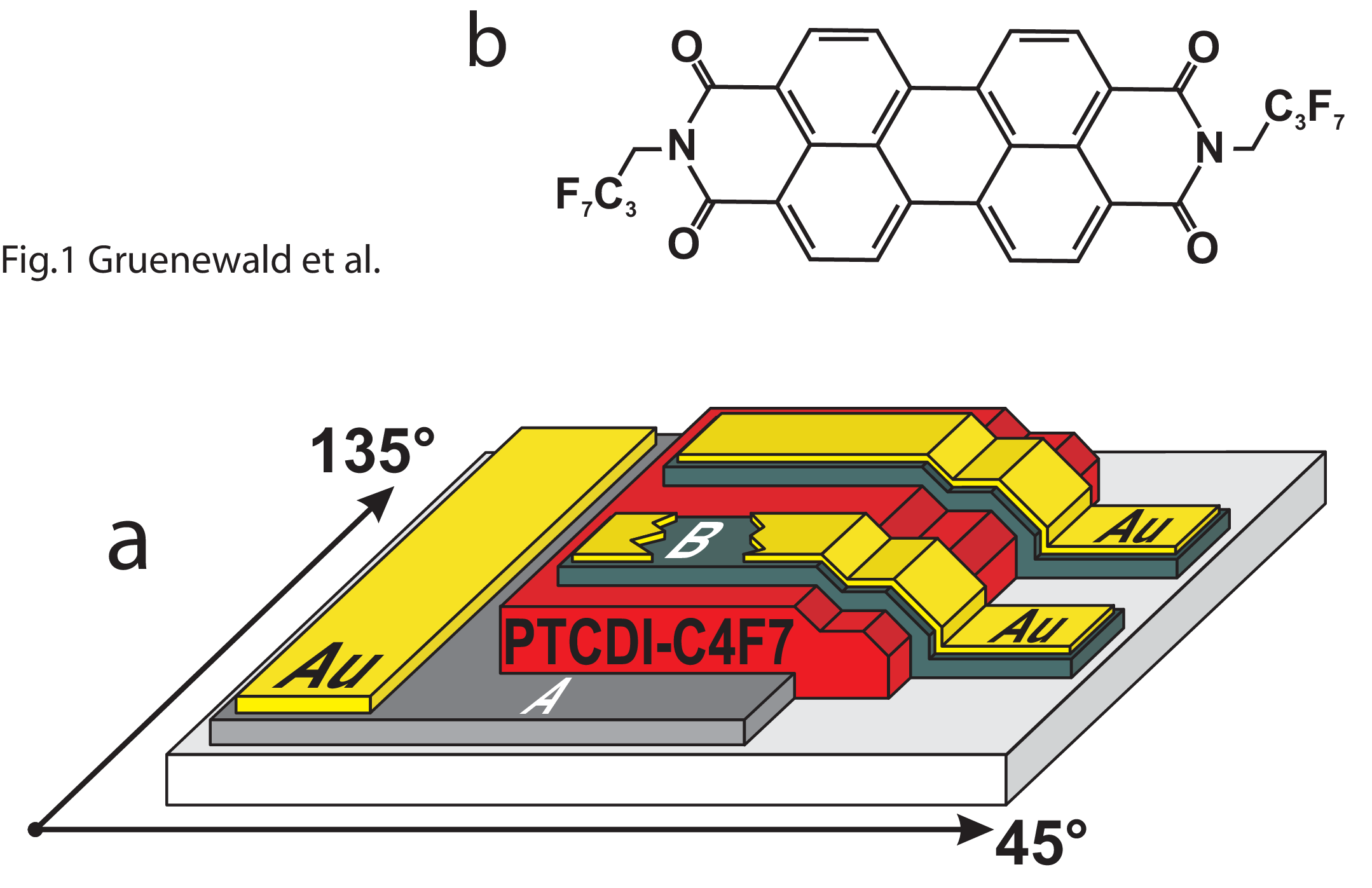}
\end{figure}

\clearpage
\begin{figure}
\includegraphics[width=18cm]{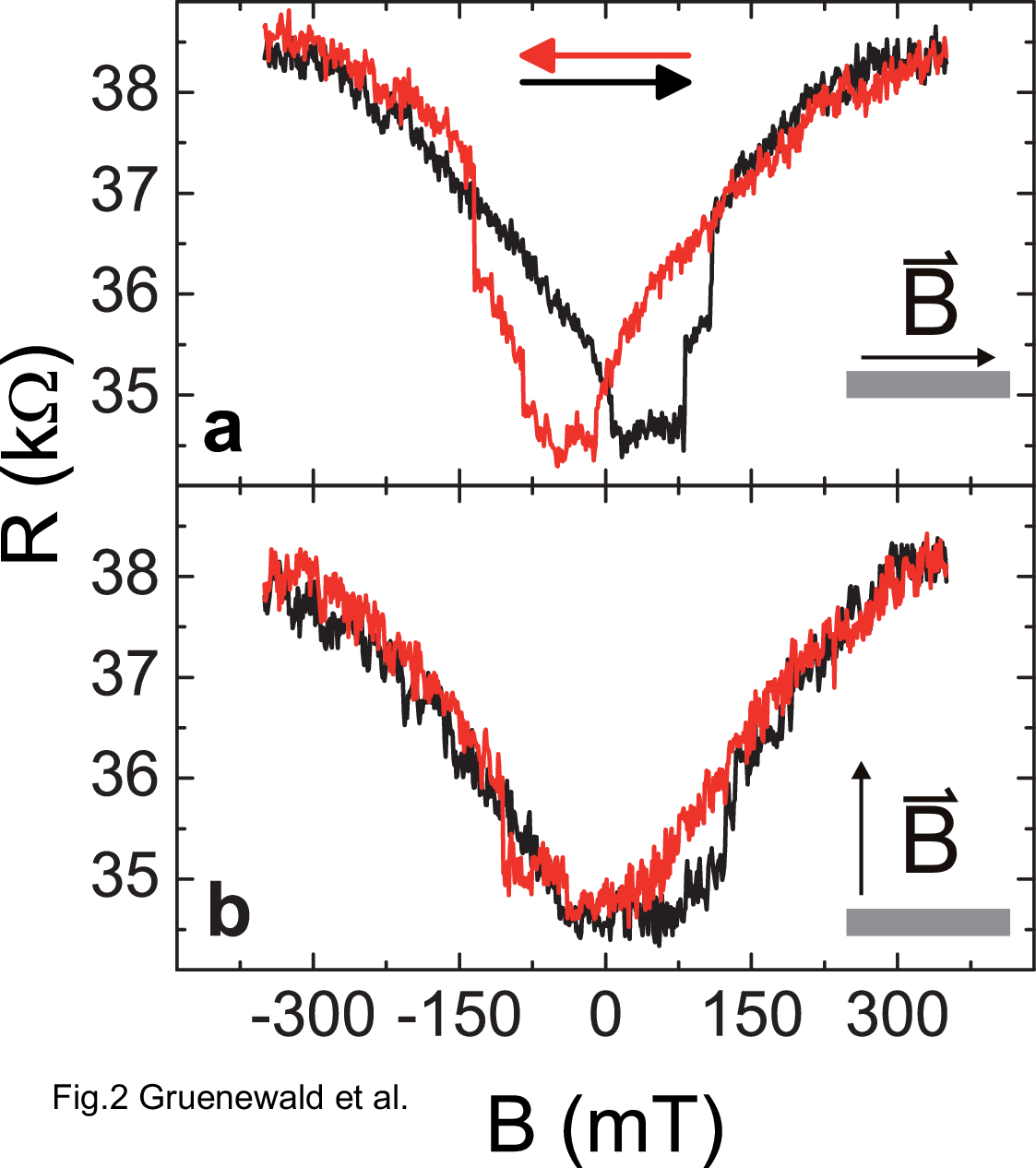}
\end{figure}

\clearpage
\begin{figure}
\includegraphics[width=18cm]{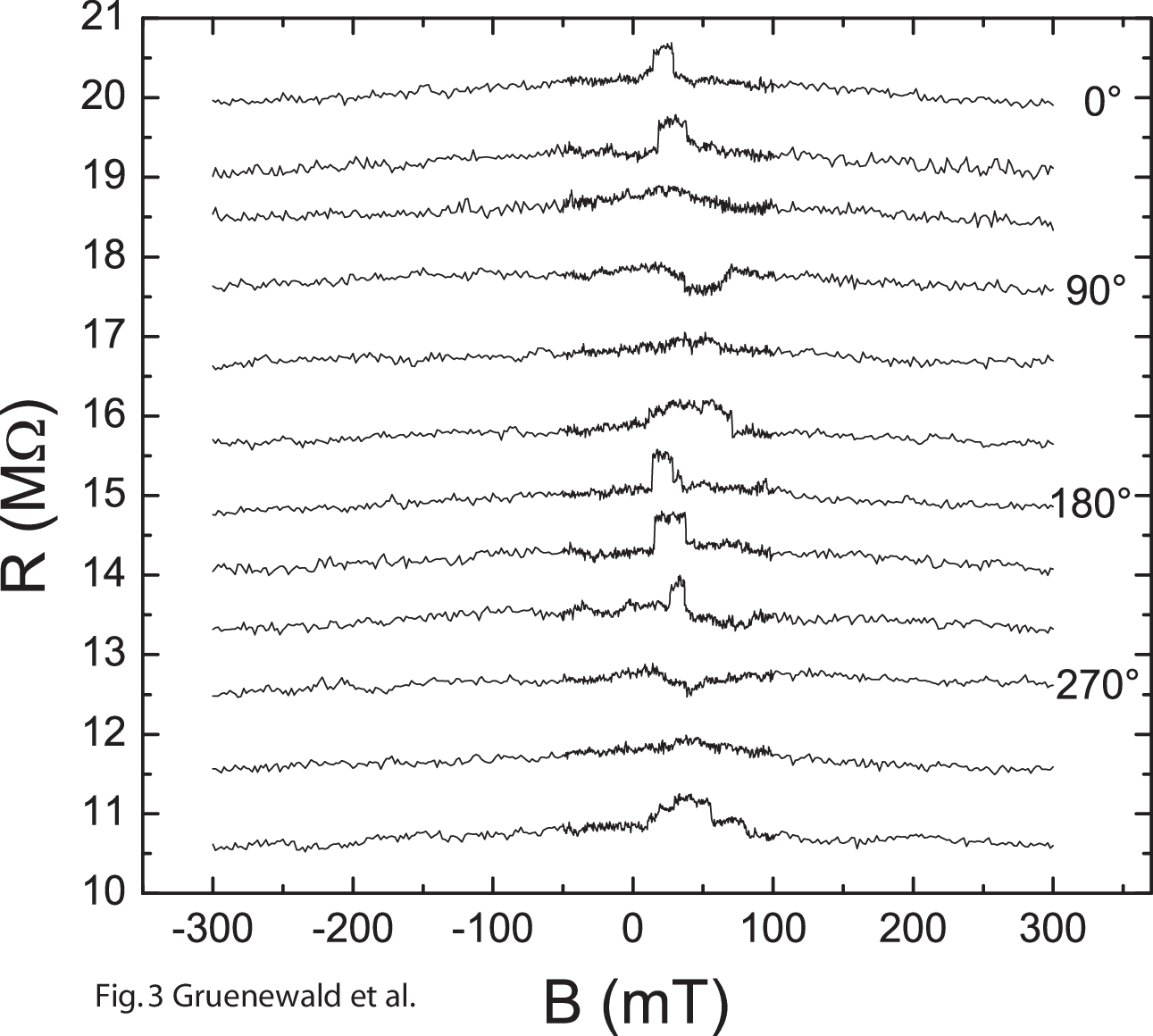}
\end{figure}

\clearpage
\begin{figure}
\includegraphics[width=18cm]{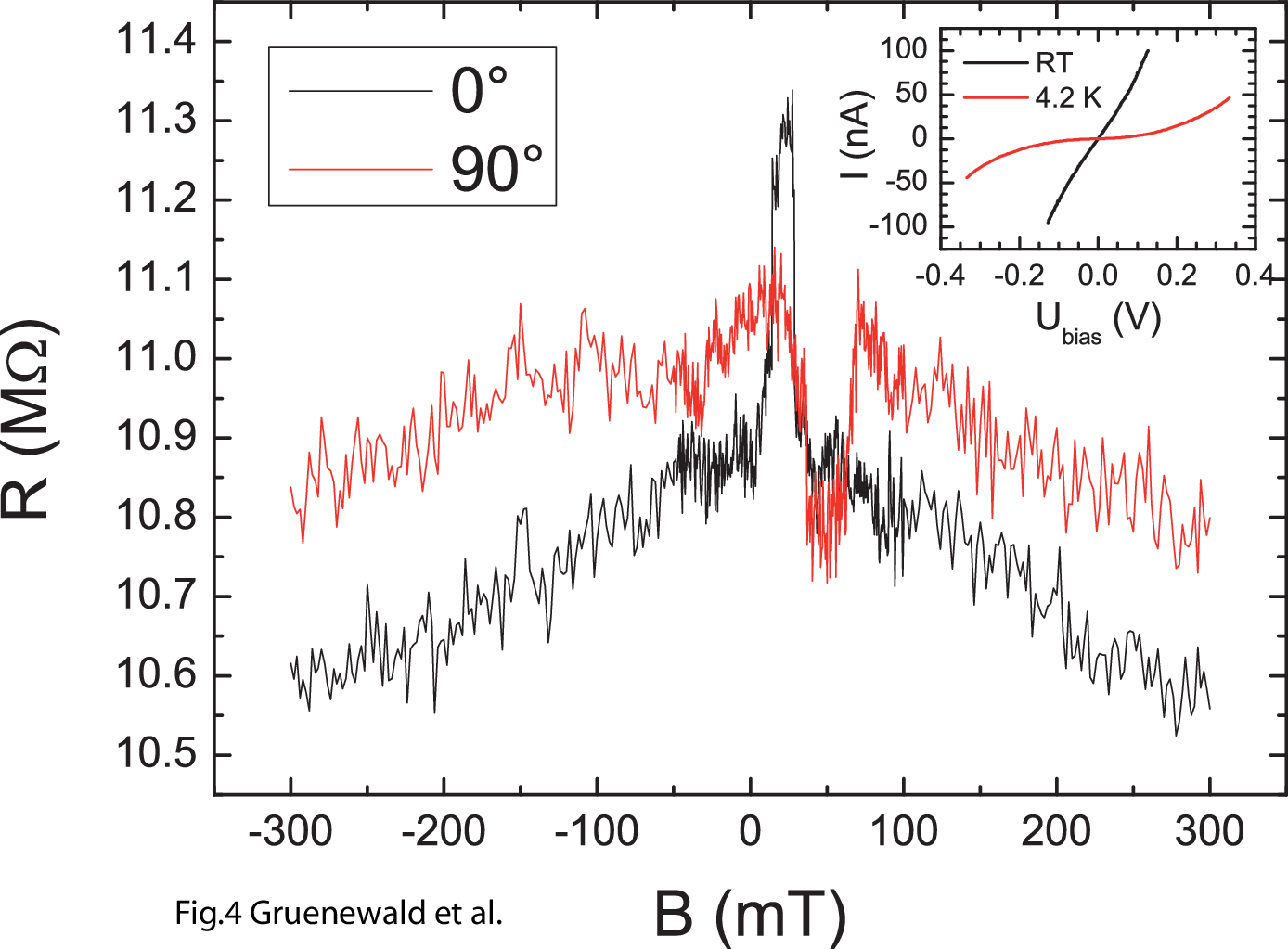}
\end{figure}

\clearpage
\begin{figure}
\includegraphics[width=18cm]{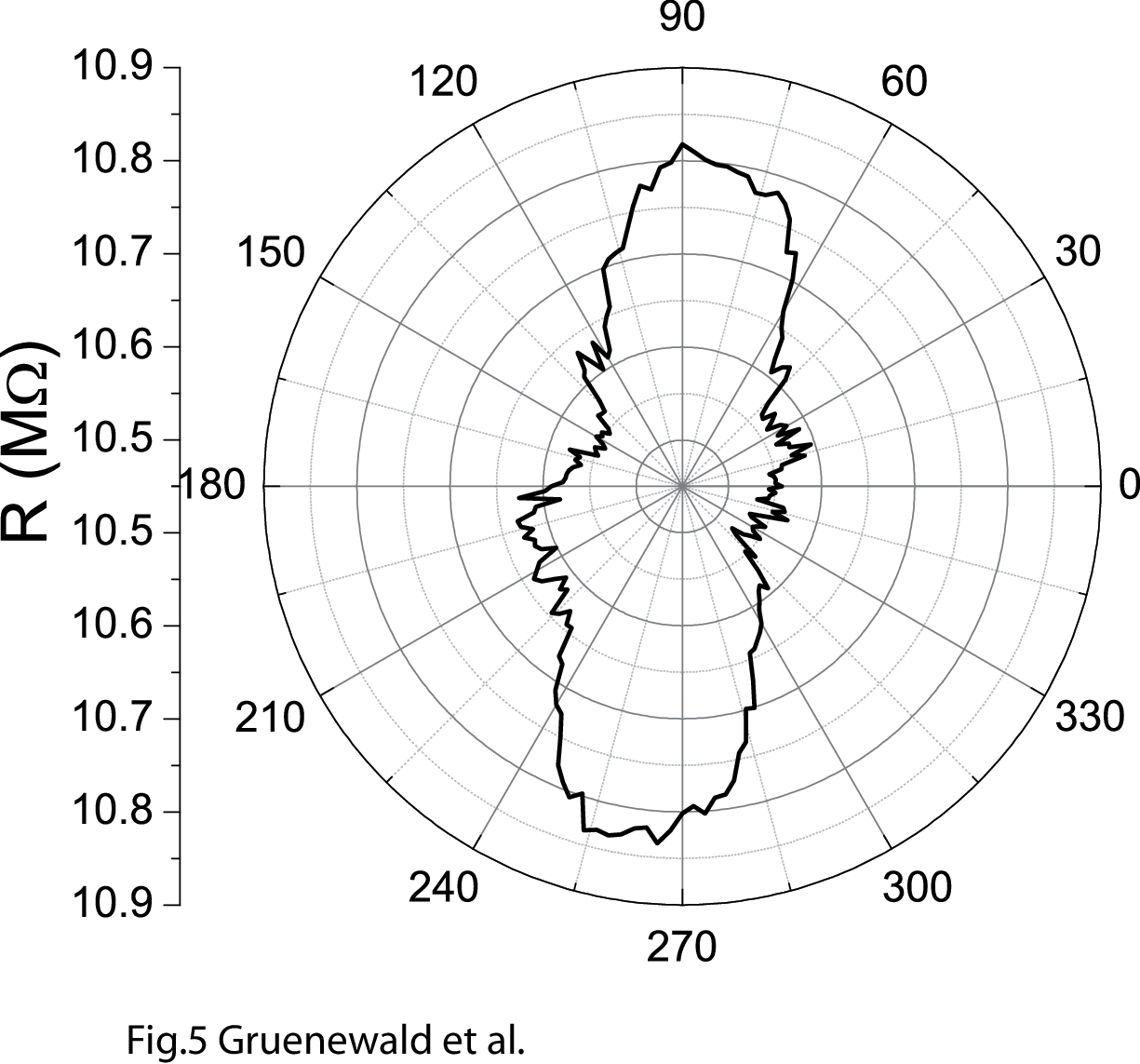}
\end{figure}


\begin{thebibliography}{99}
\bibitem{AMRnote} The in-plane magnetoresistance of the LSMO layer determined to be less than 1 $\Omega$ which is three orders of magnitude smaller than the effect observed.
\bibitem{Saturationnote} By saturation, we mean a field large enough to perfectly align the
magnetization to the direction of the external field.
\end{thebibliography}
\end{document}